# Three-Dimensional Localization of Active Aerial Targets Using a Single Terrestrial Receiver Site

Saber Kaviani and Fereidoon Behnia

*Abstract*— This paper proposes a method for the three-dimensional localization of an active aerial target by a single ground based sensor. The proposed method employs the time and frequency differences of arrival of the signal received directly from the aerial target and the signals received after being reflected from some large auxiliary terrestrial targets (pseudo-sensors) with known positions on the ground. Due to the terrestrial nature of the main and the pseudo sensors, it is impossible to solve for the target's altitude using traditional methods. The proposed method employs target motion analysis to obtain target position including its altitude with acceptable accuracy and low computational complexity. Presented simulations confirm acceptable accuracy of the proposed method in determining three dimensional position of the target despite limited number of the pseudo sensors and its low computational complexity.

## I. Introduction

THE problem of radiation source localization has drawn considerable attention in different applications such as radar [1], sonar [2], wireless sensor networks [3], and navigation systems [4]. Time difference of arrival (TDOA) and frequency difference of arrival (FDOA) are amongst the most frequently used measurements to solve for signal source location [5].

Conventional single site passive radars use the target's radiation to determine the signal direction and its parameters. In single-site localization, in addition to the angle of arrival of the signal, the target's distance is also required. The major problem of the conventional single-site passive radars is that they cannot obtain the target's distance. To solve this problem, reflection of target signal, from one or more large targets (pseudo-sensors or auxiliary targets) with known position(s), has been proposed for source localization [6]. In passive multi-sensor localization scenarios, telecommunication links are used for communication between and synchronization of central and remote stations. Using telecommunication links, however, can violate covert operation of these systems. To the contrary, in single-site localization systems, no telecommunication links are used, and signal reflection from large objects is used to simulate the role of extra sensors for target localization [7]. In these methods, the cross-correlation between the reference and the reflected signals is used to extract the needed time and frequency differences for target localization [8].

In ground based single-site localization scenarios, due to the terrestrial nature of the stations, obtaining a three-dimensional position for the target is a complex and challenging task. To address this issue, the three-dimensional localization problem is transformed to a two-dimensional problem with specified hypothetical altitude, simplifying the problem to large extent, and a closed-form expression for the target's coordinates and its two-dimensional velocity are also independently obtained. The two-dimensional coordinates of the target are derived in an explicit form using a geometric method. Two of the most frequently used geometric methods for signal source localization is the lines of position (LOP) [9, 10], and the location on the conic axis (LOCA) methods [11]. The use of the LOP and LOCA methods has many disadvantages including computational complexity, difficulty of error performance analysis, and inability to solve for parameters such as target speed. The geometric method utilized in this paper is a kind of analytic solution introduced by Fang [12]. Fang's method provides an exact solution, however, it does not make use of redundant measurements available by additional sensors to improve localization accuracy.

To solve for the target position in an n-dimensional space, we need to have at least $n + 1$ sensors, but these sensors should not be positioned in a sub-space with fewer than $n$ dimensions [13]. Here, however, all of the sensors used for localization are terrestrial, making this method ambiguous in determining the target altitude in three-dimensional space. In this paper, we use target motion analysis [14] to resolve this ambiguity, an idea inspired by the interacting multiple model (IMM) estimation method [15, 16]. In the proposed method, probabilities are assigned to each of a number of hypothetical altitudes based on a specific criterion and then the previous altitudes are updated based on the allocated probabilities. The target's two-dimensional coordinates and velocity are obtained independently in the next iteration based on the updated altitudes. This iteration continues until an acceptable level of accuracy is achieved. By extending the knowledge of existing works, the present study contributes to resolving the ambiguity of altitude in three-dimensional localization. In addition,

- We achieve a closed and explicit form for magnitude and direction of the velocity vector by assuming that the altitude of the target is known.
- We present an algorithm for obtaining the correct altitude of the target. This algorithm works as follows: first, a probability is assigned to each of a number of hypothetical altitudes, and then the following locations of the target are estimated using the Kalman and an unbiased finite impulse response (UFIR) filter. This structure operates in parallel, and the weighted estimation errors, which are calculated from the difference between the predicted position in the previous time step and the target's current position, allow optimal output selection between models attributed to the target.
- Finally, by presenting a specific criterion based on the target's current and predicted positions, we reallocate each altitude's probability and then determine the new altitude for the next iteration accordingly. The iterations continue until a desired level of accuracy is achieved.

The organization for the rest of this paper is as follows. In Section II, we find the target's position and velocity in two dimensions by assuming that the target's altitude is known. Section III presents the proposed algorithm for determining the unambiguous altitude in three-dimensions. Simulation results and subsequent discussions are given in Section IV, and finally, conclusions and a brief summary of the advantages and disadvantages of the proposed method are presented in Section V.

## II. Closed-Form Solution for the Three-Dimensional Localization at a Hypothetical Altitude

The localization method presented in this study is based on measurement of time and frequency difference of arrival of the signals collected from the central terrestrial station's two receiving channels. As shown in Fig. 1, we use the echo of the source signal reflected from the known targets (pseudo-sensors) and the source direct path signal to calculate TDOA and FDOA. In this step, the altitude of the target is assumed to be known.

In this study, only two dominant auxiliary targets with known positions are used to solve for the target position. This is because, the method used in this paper does not make use of redundant measurements provided by additional pseudo sensors and so there is no benefit in resorting to more auxiliary targets.

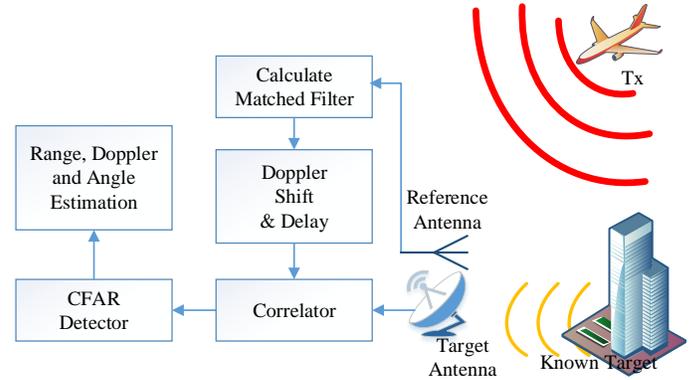

Fig. 1. Receiver block diagram of the system using pseudo sensors [18].

Fig. 1 provides an overview of the receiver structure. The signals reflected from the auxiliary targets are cross correlated with the direct path signal, and the correct Doppler and delay values are extracted using the constant false alarm rate (CFAR) block. The reason for using the CFAR block is to provide some degree of immunity to unwanted/uncorrelated high power signals. In addition, by choosing the suitable type and setting proper coefficients for the CFAR, the effects of unequal gain of the receiver in the frequency domain can be partially compensated. Fig.2 illustrates how to use targets with known positions in determining the target's position.

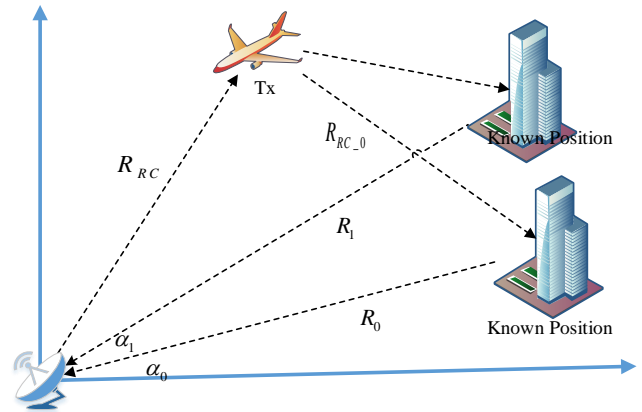

Fig. 2. Exemplar placement of the target, the main station and the pseudo sensors

### A. Closed-form solution for TDOA/FDOA

In this subsection, we assume that the target altitude is known. Using Fang's method and TDOA equations, we calculate the target position in two dimensions and the target altitude is considered as a free parameter. Fang's method has the following features

- Exact solution, no approximations
- Suitable for ground-based sensors
- Number of measurements is equal to the number of unknowns
- No noise consideration
- Cannot use extra measurements (sensors)

In this method, three sensors and two independent measurements are sufficient for two-dimensional positioning, and the altitude of the target is assumed to be known. After obtaining the two-dimensional coordinates of the target, we convert these coordinates into the spherical distance, the azimuth angle, and the elevation angle as below

$$\begin{cases} R_i = \sqrt{x_i^2 + y_i^2 + h_i^2} \\ \alpha_{azimuth} = tag^{-1}(\frac{y_i}{x_i}) \\ \alpha_{elevation} = tag^{-1}(\frac{h_i}{\sqrt{x_i^2 + y_i^2}}) \end{cases} \quad (1)$$

In the next step, the target speed is obtained explicitly in two dimensions in a closed-form. To do end, first of all, we examine the vector form of FDOA equation and then derive its scalar and nonlinear equivalent, and then, we obtain the target position in three dimensions for a predetermined altitude. The linear and vector-form of the equation is

$$\dot{s}_i^T x + \dot{r}_{io} r_0 + s_i^T \dot{x} + r_{io} \dot{r}_0 = \dot{s}_i^T s_i^T - r_{io} \dot{r}_{io} \quad (2)$$

where $s_i$ is the position of the i-th sensor, $x$ indicates the position of the target, $r_i$ is the value of the time difference of arrival, $\dot{r}_i$ is the value of the frequency difference of arrival and $r_0$ represents the distance between the active aerial target and the origin. Furthermore, $\dot{r}_0$ is the derivative of $r_0$. The scalar-form of (2) can be written as

$$\begin{aligned} & s_i v \cos(\theta_v - \theta_{s_i}) + \dot{r}_{io} R + \\ & r_{io} v \cos(\theta_v - \alpha_{azimuth}) \cos(\alpha_{elevation}) = 0 \\ & \begin{cases} |\dot{x}| = v \\ |x - s_0| = r_0 = R \end{cases} \end{aligned} \quad (3)$$

Where $\dot{s}_i = 0$ is assumed considering that fact the main and the pseudo sensors are motionless. Since there are two auxiliary targets, we have a two equations two unknowns system of equations for $v$ and $\theta_v$.

A traditional way to solve this system of equations is to use a grid search, which is a method with high computational complexity, especially when the search area is three-dimensional and/or high accuracy for localization is required. Therefore, another method is proposed to solve this problem. In this method, we expand the previous relation by using trigonometric identities, and then obtain the variables explicitly and independently.

As shown in Appendix, we arrive at the equations for $v$ and $\theta_v$ as outlined in the following. Considering the normal flight scenario (no abrupt changes in altitude), we assume $\theta_v$ to be two-dimensional. We track altitude changes for the aerial platform in subsequent iterations of the algorithm.

$$\begin{aligned} v^2 &= \frac{(r_{10}\dot{r}_{10} - \dot{r}_{10}R)^2(A_2^2 + B_2^2)}{(A_1B_2 - A_2B_1)^2} \\ &+ \frac{(r_{20}\dot{r}_{20} - \dot{r}_{20}R)^2(B_1B_2 + A_1A_2)^2}{(A_2^2 + B_2^2)(A_1B_2 - A_2B_1)^2} \\ &+ 2 \times \frac{(r_{10}\dot{r}_{10} - \dot{r}_{10}R)(r_{20}\dot{r}_{20} - \dot{r}_{20}R)(B_1B_2 + A_1A_2)}{(A_1B_2 - A_2B_1)^2} \\ &+ \frac{(r_{20}\dot{r}_{20} - \dot{r}_{20}R)^2}{A_2^2 + B_2^2} \end{aligned} \quad (4)$$

To simplify, we use the following substitutions in parts of (4).

$$\begin{cases} A_i = s_i \cos(\theta_{s_i}) + r_{i0} \cos(\alpha_{azimuth}) \cos(\alpha_{elevation}) \\ B_i = s_i \sin(\theta_{s_i}) + r_{i0} \sin(\alpha_{azimuth}) \cos(\alpha_{elevation}) \end{cases} \quad (5)$$

After determining the exact value of $v$, we calculate the exact value of the velocity vector angle $\theta_v$.

$$\theta_v = sin^{-1}(\frac{r_{10}\dot{r}_{10} - \dot{r}_{10}R}{v \times \sqrt{A_1^2 + B_1^2}}) - tag^{-1}(\frac{A_1}{B_1}) \quad (6)$$

At the beginning of this subsection, target coordinates were calculated for each hypothetical altitude, and at the end of this subsection, the velocity vector's magnitude and direction were obtained explicitly and independently. In the next section, we will use a method to solve the problem of target altitude determination.

### III. SOLVING FOR TARGET ALTITUDE BY MOTION ANALYSIS

This method, which is based on motion analysis of aerial targets, solves for the target altitude iteratively [19, 20]. As we know, Fang's method suffers from considerable levels of error in noisy conditions since it cannot use additional measurements (extra sensors) and cannot solve the problem in overdetermined conditions. Moreover, one cannot localize the target in three-dimensions, using three sensors. Thus, we use motion analysis to increase the localization accuracy of the mentioned method and to resolve the ambiguity of the target altitude.

#### A. Motion Analysis and Increasing Localization Accuracy

Due to the iterative nature of the algorithm presented in this paper, positioning errors in successive iterations can lead to non-convergence of the algorithm. For this reason and in order to increase the localization accuracy in the first step, we exploit integration of Kalman and UFIR filters to take advantage of accuracy and robustness provided by the two [19]. These operations are performed by the adaptive interacting multiple model (IMM) block which estimates the optimal state vector and covariance matrix [19], as shown in the block diagram of Fig. 3.

Multiple known targets (pseudo-sensors) are fixed in predetermined locations, and one central terrestrial station with receiver channels is used for signal processing to extract TDOA and FDOA values and determine the position and the velocity of the moving aerial target. The prediction of the target state vector $X_k$ at the time step $t_k$ as a function of its previous state $t_k$ is given by

$$X_{k|k-1} = FX_{k-1|k-1} + W_k \tag{7}$$

Where $W_k$ is the zero mean process noise, (not necessarily Gaussian) with covariance matrix $Q_k$, and $X_k$ indicates the aerial platform's state which consists of three dimensional position and velocity components of the target as below

$$X_j = [x, y, h_j, V_x, V_y, 0] \tag{8}$$

Assuming the interval between successive measurements to be short, a constant velocity motion model is considered to estimate the next position of the target. The advantage of using FDOA data in this section is that in the Kalman and UFIR filters' update stage, the magnitude and direction of the velocity vector of the aerial moving target are also corrected. In this case, the measurement vector $Z(k)$ becomes $[x(k)\ y(k)\ h_i(k)\ Vx(k)\ Vy(k)\ V_z(k)]^T$ instead of $[x(k)\ y(k)\ h_i(k)]^T$, and targets with more complex motion models can be localized using the same assumptions. Therefore, the Kalman and the UFIR filters are applied to the position and speed of the target with a constant speed transfer matrix. The state-transition matrix for this system is.

$$F = \begin{bmatrix} I_{3\times 3} & \Delta t \times I_{3\times 3} \\ 0_{3\times 3} & I_{3\times 3} \end{bmatrix} \tag{9}$$

Here the notations $I_{n\times n}$ and $0_{n\times n}$ indicate $n \times n$ identity matrix and $n \times n$ zero matrix, respectively. $\Delta t$ refers to the time step and $W$ is the white Gaussian noise process.

The observation equation representing $Z_k$ from TDOA/FDOA is given as

$$Z_k = HX_{k|k-1} + v_k \tag{10}$$

Where $v_k$ is the white Gaussian noise vector, and $H$ is

$$H = I_{6\times 6} \tag{11}$$

The variances of TDOA and FDOA measurement are denoted by $\sigma_T^2$ and $\sigma_F^2$, respectively.

### B. Resolving the Ambiguity of the Target Altitude

As mentioned earlier, we first use the adaptive IMM estimator block to increase the positioning accuracy at any hypothetical altitude, and then, as shown in Fig. 3, the outputs of this block is used along with the prediction of the optimal motion model to calculate for the allocated probability for each altitude.

This method is applied in two levels. After extracting the adaptive IMM estimator block outputs, the prediction for the optimal motion model is made by Time Update block as

$$X_{k|k-1} = FX_{k-1|k-1}$$
$$P_{k|k-1} = FP_{k-1|k-1}F^T + Q_k \tag{12}$$

The function $\Lambda_j(k)$ is defined as the probability density function of the estimated value $H_j X_j^-(k)$, which is assumed to be normal with mean $Z(k)$ and covariance $S_j(k)$. This density function is a normal distribution around the value estimated by the prediction model, as inspired by the method described in [18, 19].

$$\begin{cases} S_j(k) = H_j . P_j^-(k) . H_j^T + R_j \\ \tilde{y}_j(k) = Z(k) - H_j X_j^-(k) \\ d_j^2(k) = \tilde{y}_j(k)^T . S_j(k)^{-1} . \tilde{y}_j(k) \\ \Lambda_j(k) = \dfrac{\exp(\dfrac{-d_j^2(k)}{2})}{\sqrt{|2\pi S_j(k)|}} \end{cases} \tag{13}$$

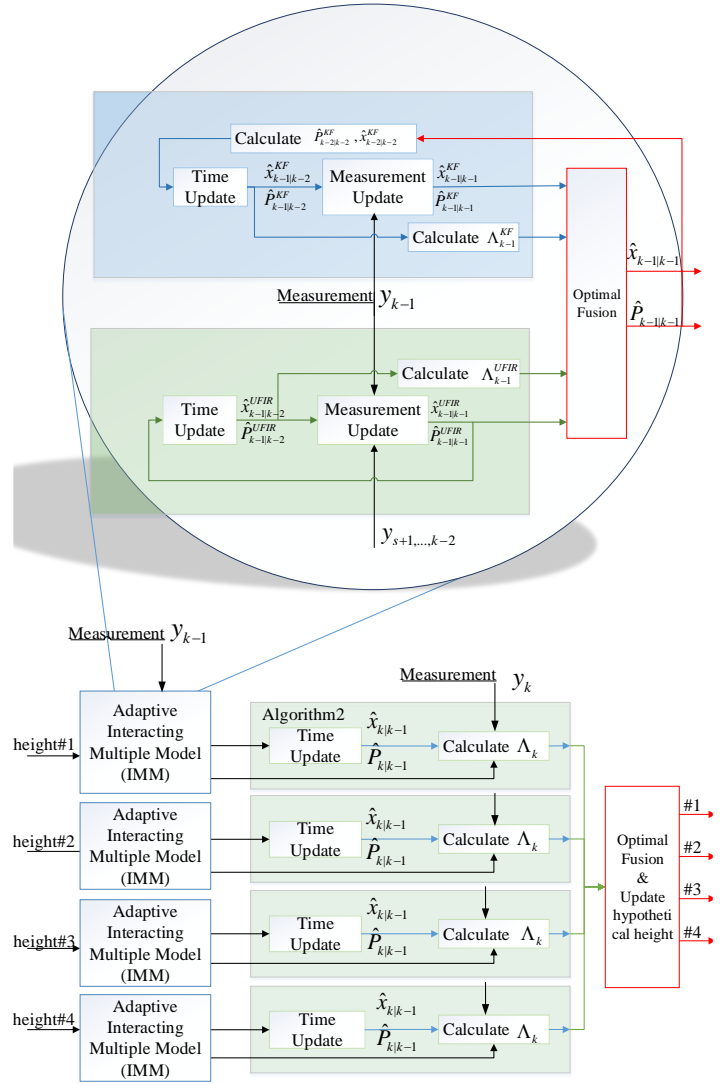

Fig. 3. Block diagram of the proposed method.

In (13), the indices represent the state vectors for each hypothetical altitude (j=1, 2, 3, 4). The main part of the proposed algorithm is the Optimal Fusion and Update of the hypothetical altitude. In this block, the updated probability values assigned to each altitude $u_j(k)$ are calculated using $\Lambda_j(k)$ and the probabilities assigned to each altitude in the previous iteration. In the first iteration, equal probabilities are considered for the four hypothetical altitudes of the aerial target (for example [0, 5km, 10km, 15km]).

$$u_j(k) = \frac{\Lambda_j(k)u_j(k-1)}{\sum_{i=1}^{4}\Lambda_i(k)u_j(k-1)} \quad (14)$$

Then, considering the Markov model for the transfer probabilities from one altitude to another one ($\rho_{ij}$) Bayes' law is used to calculate the transfer probability from one state to another one as below

$$U_{ij}(k) = \frac{\rho_{ij}u_i(k)}{\sum_{l=1}^{4}\rho_{lj}u_l(k)} \quad (15)$$

In (15), $U_{ij}(k)$ is the conditional probability of the $j$th state in time step $k$, provided the target was in the $i$th state at the time step $k-1$. The state here refers to the motion profile at the altitudes assumed for the aerial target. The $X_j$ state vector was defined previously. In fact, as can be seen in the definition of $X_j$, we have not considered the parameters of altitude and speed in the direction of altitude because the altitude is updated in each iteration and is known as the state tag. In the next step, the new inputs of the motion model are calculated according to the conditional probabilities obtained in the previous step.

$$h_j^0(k) = \sum_{i=1}^{4} h_i^+(k)U_{ij}(k) \quad (16)$$

The parameter $h_i^+(k)$ in (16) is equal to the updated value of the system state tag (altitude of target) in the previous iteration, which can be the initial value or the output obtained from the previous iteration. As in section *II*, the value $h_j^0(k-1)$ is used to calculate the position and speed of the target.

Finally, according to the values of h(k) calculated in (2) and using the method proposed in section II, the two-dimensional coordinates and the velocity related to the new values of the updated altitudes h(k) are obtained. The pseudocode of the proposed iterative method can be seen below in Algorithm 1. In this pseudocode, first the position of the target and its velocity vector are obtained, and then using the algorithm presented in the previous section, the hypothetical altitudes of the target are updated. This is repeated until the altitude difference between two iterations becomes less than a certain value. This value, which depends on the SNR of the received signal and the target position, is denoted by $\varepsilon_0$.

**Algorithm 1:** Solve the Problem of Target Altitude

Initialize total each variable
Hypothetical initial value of altitude = 0, 5km, 10km, 15km
$h_i$ = Hypothetical initial value of altitude // i=1,2,3,4
$u_i$ = 0.25 // Assign initial probability to any altitude
Data: $x, y, V_x, V_y$
Result: $x, y, z, V_x, V_y$
Begin
1  **while** $|h_i - h_j| > \varepsilon_0$ // $\forall$ i , j  ;  i≠j
   // Step 1. Acquire target position and speed (section II)
2  $X_{k-1|k-1} = [x. y. V_x. V_y]$
   // Step 2. Proposed algorithm (section III)
3  $X_{k|k-1}^i = predict(X_{k-1|k-1}^i)$ // UFIR or Kalman filter
4  $X_{k|k}^i = update(X_{k-1|k-1}^i, LinearModel)$ // Optimum filter
5  $d_i = Z - X_{k|k}^i$ // i=1,2,3,4
   // Probability density function $\Lambda_i$ According to $d_i$
6  $\Lambda_i(k)$ is obtained from $Eq. (13)$.
   // Assign probability to any altitude
7  $u_i(k)$ is obtained from $Eq. (14)$.
8  $h_{maximum\_likelihood} = h\left(arg\max_{1 \leq i \leq 4}(u_i)\right)$
9  The values of $h_i^+$ are updated **using** $Eq. (16)$.
10 End **while**
11 $Z_{Target} = h_{maximum\_likelihood}$
12 End begin

## IV. SIMULATION RESULTS

In this section, numerical simulations are employed to evaluate the performance of the proposed method. For this purpose we assume the following placement for the main and the pseudo sensors. The main sensor is placed at the origin, and the two pseudo sensors are assumed to be placed at (10, 0, 0)km and (10, 10, 0)km respectively. The aerial target is assumed to be situated at (20, 5, $h_i$)km, and simulations are performed for different values of $h_i$. Also, target speed is set as $v_x = v_y = 70 \frac{m}{s}$.

We consider the following form for the matrix $\rho$, so that at each new step, the probability of staying at the same altitude is the same as that of moving to any of the adjacent altitudes for the target. The effect of considering these values for the matrix can be seen in (15).

$$\rho = \begin{bmatrix} 0.5 & 0.5 & 0 & 0 \\ 0.33 & 0.33 & 0.33 & 0 \\ 0 & 0.33 & 0.33 & 0.33 \\ 0 & 0 & 0.5 & 0.5 \end{bmatrix}$$

The initial values of probabilities are also considered as $u_i = \frac{1}{M}$, in which $M = max(i) = 4$.

First, the target is assumed to be at the altitude of 10 km, and the result is plotted for $u_i$ after repeating all of the above steps once. The result is shown in Fig. 4.

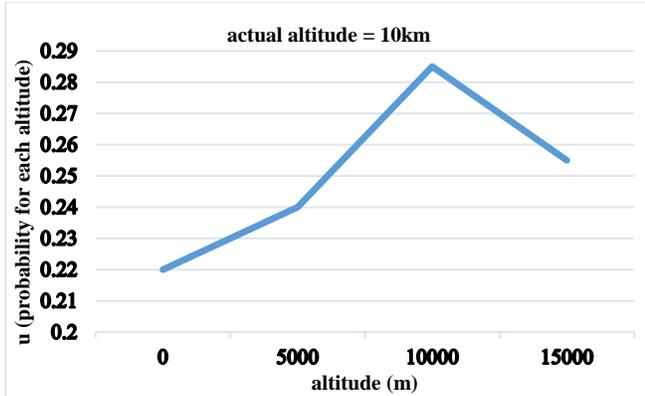

Fig. 4. How to distribute the probability of flying at any altitude after one repetition.

As shown in Fig. 4, the target is more likely to be at an altitude of 10km than any other altitude. The algorithm is continued to check the achievable accuracy and improve the resolution of the target altitude report. The result is shown in Fig. 5 after nine repetitions of all steps of the algorithm.

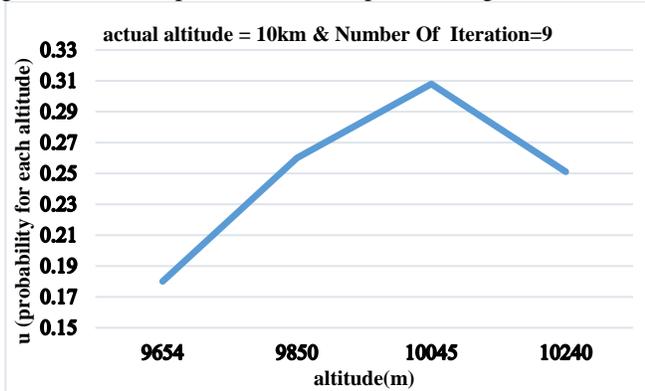

Fig. 5. How to distribute the probability of flying at any altitude after nine repetitions.

As can be seen, the candidate altitude values [0, 5, 10, 15]km in the first iteration (as shown in Fig. 4) change to [9654, 9850, 10045, 10240]m after 9 repetitions in Fig. 5 being more close to the actual altitude of 1okm.

In order to check the results for altitudes other than the default altitudes of [0, 5, 10, 15]km, we assume that the target flies at an altitude of 8.5km, then the result of the algorithm after 9 repetitions is as shown in Fig. 6, which clearly indicates that the probability of being at the altitude of 8560m is more than any other altitude, and this altitude is reported as the altitude of the target by the algorithm. Note that this result is sufficiently close to the actual altitude of 8500m.

Fig. 7 shows how the algorithm converges to the altitude of the target at altitudes of 10km and 8.5km. In this Figure and in each iteration, the altitude with the highest probability of $u_i$ is selected (14). As can be seen, in both cases, after some transient states, the altitudes converge from an initial default value to the correct altitude.

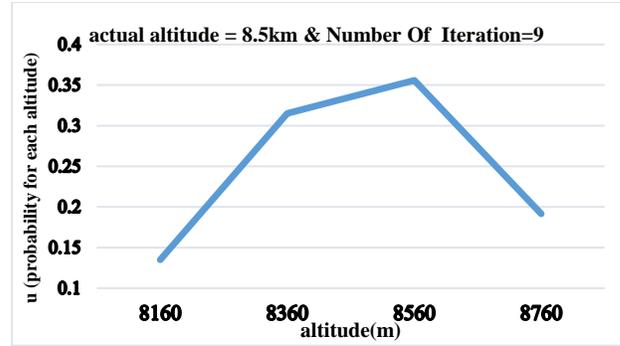

Fig. 6. How to distribution the probability of flying at any altitude after nine repetitions.

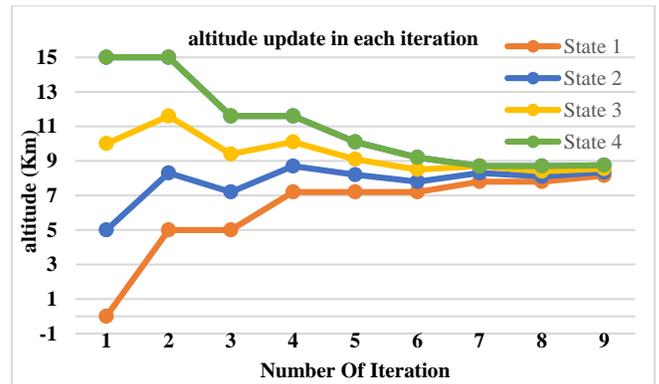

Fig. 7. Updated values for altitude in the fifth stage of the proposed method (eqn.14)

Table 1 shows the probability distribution among the four states in each iteration.

TABLE I
SIMULATION RESULTS IN EACH ITERATION

| Number of iteration | | $i = 1$ | $i = 2$ | $i = 3$ | $i = 4$ |
|---|---|---|---|---|---|
| 0 | $h_i(km)$ | 0 | 5 | 10 | 15 |
|   | $u_i$ | 0.25 | 0.25 | 0.25 | 0.25 |
| 1 | $h_i(km)$ | 0 | 5 | 10 | 15 |
|   | $u_i$ | 0.262 | 0.298 | 0.381 | 0.121 |
| 2 | $h_i(km)$ | 5 | 8.3 | 11.6 | 15 |
|   | $u_i$ | 0.291 | 0.321 | 0.269 | 0.118 |
| 3 | $h_i(km)$ | 5 | 7.2 | 9.4 | 11.6 |
|   | $u_i$ | 0.207 | 0.325 | 0.335 | 0.132 |
| 4 | $h_i(km)$ | 7.2 | 8.7 | 10.1 | 11.6 |
|   | $u_i$ | 0.296 | 0.326 | 0.256 | 0.12 |
| 5 | $h_i(km)$ | 7.2 | 8.2 | 9.1 | 10.1 |
|   | $u_i$ | 0.231 | 0.34 | 0.299 | 0.128 |
| 6 | $h_i(km)$ | 7.2 | 7.8 | 8.5 | 9.2 |
|   | $u_i$ | 0.13 | 0.274 | 0.354 | 0.244 |
| 7 | $h_i(km)$ | 7.8 | 8.3 | 8.7 | 8.7 |
|   | $u_i$ | 0.188 | 0.349 | 0.325 | 0.136 |
| 8 | $h_i(km)$ | 7.8 | 8.1 | 8.4 | 8.7 |
|   | $u_i$ | 0.125 | 0.255 | 0.339 | 0.28 |
| 9 | $h_i(km)$ | 8.15 | 8.35 | 8.55 | 8.75 |
|   | $u_i$ | 0.137 | 0.315 | 0.357 | 0.19 |

In the following, we compare the distance detection error in this study with that of the conventional two-step and linear methods[21]. In these methods, five receivers are used for passive three-dimensional localization of a target using the time and frequency differences of the received signals. In contrast, in the method presented in this study a single receiver along with two pseudo sensors are used. Therefore, to compare the conventional methods with the method presented in this study, only three receivers are considered. Our method is very advantageous to other methods as long as we are limited to three terrestrial receivers. The arrangement of the three receivers to simulate the mentioned methods is such that the first (main) receiver is placed at the origin and the two pseudo sensors are placed in two arbitrary positions. TDOA and FDOA measurements are derived adding Gaussian noise with a mean of zero and diagonal covariance matrices to each parameter's exact value. The diagonal elements of the covariance matrices are considered as $\sigma_T^2 = \sigma^2$ and $\sigma_F^2 = 0.1\sigma^2$ for TDOA and FDOA measurements respectively and the noise terms of TDOA and FDOA measurements are assumed to be uncorrelated. The position and speed of the target are considered as before with an altitude of 8.5km. In the methods that are to be compared with this study, as long as just three receivers are used the aerial target coordinates will be measured in two dimensions only, ignoring the altitude and solving for target distance in two dimensions. Failure to consider the target altitude adds a fixed error to the calculations, which is

$$Error_{Fixed} = \sqrt{x^2 + y^2 + h^2} - \sqrt{x^2 + y^2}$$
$$= \sqrt{20km^2 + 5km^2 + 8.5km^2} - \sqrt{20km^2 + 5km^2}$$
$$= 1.68km$$

After 1000 runs of the simulation for each value of noise variance in all three methods, the result of localization accuracy is derived as shown in fig.8.

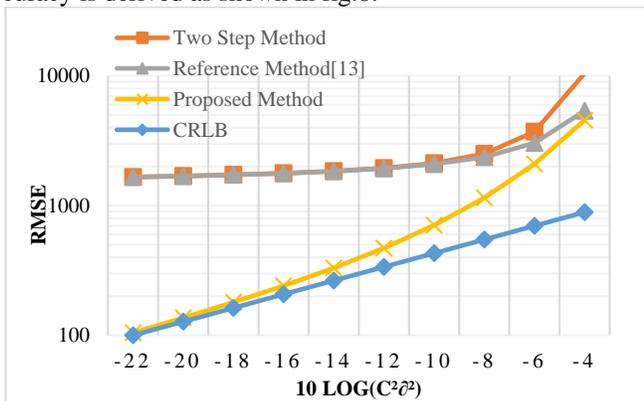

Fig. 8. Comparison of different methods with the method presented in this method

As shown in this figure, the proposed method performs considerably better compared to the two competing methods.

## V. Conclusion

In this study we considered a single ground based sensor along with two large terrestrial reflectors (pseudo-sensors) to obtain the three dimensional position of an aerial target. To this end, a closed-form solution was presented for the two dimensional position of the target based on TDOA and FDOA equations, and motion analysis was used to derive target altitude despite the fact that all of the main and pseudo stations were placed on the ground.

In addition to being able to derive the three dimensional location of an aerial target based on just one main and few pseudos' ground based sensors, the presented algorithm also has low computational complexity making it attractive for practical applications.

Presented simulations validated performance of the algorithm and showed its superiority to conventional methods in determining position of aerial targets using ground based sensors.

### Appendix

This appendix provides a closed and explicit formula for obtaining the magnitude and direction of the velocity vector of the aerial target using (2). To elaborate, first we rewrite equation (2) here

$$\dot{s}_i^T x + \dot{r}_{io} r_0 + s_i^T \dot{x} + r_{io} \dot{r}_0 = \dot{s}_i^T s_i^T - r_{io} \dot{r}_{io} \quad (17)$$

Due to the immobility of sensors, the terms $\dot{s}_i^T x$ and $\dot{s}_i^T s_i^T$ are considered zero. As mentioned earlier, $r_{io}$ represents the input time difference, $\dot{r}_{io}$ represents the input frequency difference, and $r_0$ represents the magnitude of the target position vector, which measures the magnitude ($R_i$) and direction ($\alpha_i$) in (1). The parameter $\dot{r}_0$ is the derivative of the magnitude of the target position vector ($R_i$), which must be written in terms of magnitude and the direction of the velocity vector ($v, \theta_v$). Let's first write the formula for derivative of a vector with respect to time

$$\frac{\partial |\vec{a}|}{\partial t} = \frac{\partial (\vec{a}.\vec{a})^{(1/2)}}{\partial t} = \frac{\vec{a}.\frac{\partial \vec{a}}{\partial t}}{|\vec{a}|}$$
$$= |\frac{\partial \vec{a}}{\partial t}| \times cos(\theta_{\frac{\partial \vec{a}}{\partial t}} - \theta_{\vec{a}}) \quad (18)$$

In similarly to (18), we write the following relation for the parameter $\dot{r}_0$ or $\dot{R}_i$

$$\dot{r}_0 = \dot{R}_i = \frac{\partial |\vec{R}|}{\partial t} = |\frac{\partial \vec{R}}{\partial t}| \times cos(\theta_{\frac{\partial \vec{R}}{\partial t}} - \theta_{\vec{R}}) \quad (19)$$
$$\Rightarrow \dot{r}_0 = v \times cos(\theta_v - \alpha_{azimuth}) cos(\alpha_{elevation})$$

The following equation is used to calculate the spherical angle in (19).

$$\cos(\alpha_{3\text{-D}}) = \cos(\alpha_{azimuth}) \times \cos(\alpha_{elevation}) \quad (20)$$

Finally, we obtain the scalar-form of (17), as follows

$$s_i v \cos(\theta_v - \theta_{s_i}) + \dot{r}_{io} R + \ldots \\ r_{io} v \cos(\theta_v - \alpha_{az}) \cos(\alpha_{el}) + r_{io} \dot{r}_{io} = 0 \quad (21)$$

By simplifying (21), we achieve an explicit and independent form for each unknown. To do this, the sine and cosine functions of (21) are expanded

$$s_i v \cos(\theta_v) \cos(\theta_{s_i}) + r_{io} v \cos(\theta_v) \cos(\alpha_{azi}) \cos(\alpha_{el}) + \ldots \\ s_i v \sin(\theta_v) \sin(\theta_{s_i}) + r_{io} v \sin(\theta_v) \sin(\alpha_{az}) \cos(\alpha_{el}) \\ = r_{io} \dot{r}_{io} - \dot{r}_{io} R \quad (22)$$

Then (22) is rearranged, and the two sides are divided by $v$ to yield

$$s_i \cos(\theta_v) \cos(\theta_{s_i}) + r_{io} \cos(\theta_v) \cos(\alpha_{az}) \cos(\alpha_{el}) + \ldots \\ s_i \sin(\theta_v) \sin(\theta_{s_i}) + r_{io} \sin(\theta_v) \sin(\alpha_{az}) \cos(\alpha_{el}) \\ = \frac{r_{io} \dot{r}_{io} - \dot{r}_{io} R}{v} \quad (23)$$

For brevity, parts of (23) are substituted as follows

$$\begin{cases} A_i = s_i \cos(\theta_{s_i}) + r_{i0} \cos(\alpha_{azi}) \cos(\alpha_{ele}) \\ B_i = s_i \sin(\theta_{s_i}) + r_{i0} \sin(\alpha_{azi}) \cos(\alpha_{ele}) \end{cases} \quad (24)$$

By using (24) in (23) and simplifying the result, the following equations are formed for $i = 1\ \&\ 2$.

$$\begin{cases} \sin(\theta_v + tag^{-1}(\frac{A_1}{B_1})) = \frac{r_1 \dot{r}_{1o} - \dot{r}_{1o} R}{v \sqrt{A_1^2 + B_1^2}} & (I) \\ \sin(\theta_v + tag^{-1}(\frac{A_2}{B_2})) = \frac{r_2 \dot{r}_{2o} - \dot{r}_{2o} R}{v \sqrt{A_2^2 + B_2^2}} & (II) \end{cases} \quad (25)$$

Equation (I) of (25) is rewritten as

$$\sin\left(\theta_v + tag^{-1}(\frac{A_2}{B_2}) + \left(tag^{-1}(\frac{A_1}{B_1}) - tag^{-1}(\frac{A_2}{B_2})\right)\right) = \\ \frac{r_1 \dot{r}_{1o} - \dot{r}_{1o} R}{v \sqrt{A_1^2 + B_1^2}} \quad (26)$$

After expanding the sine function and substitution of part (II) of (25), we arrive at the following equation, the only variable of which is $v$:

$$\frac{r_2 \dot{r}_{2o} - \dot{r}_{2o} R}{v \sqrt{A_2^2 + B_2^2}} \times \cos\left(tag^{-1}(\frac{A_1}{B_1}) - tag^{-1}(\frac{A_2}{B_2})\right) + \\ \sqrt{1 - \frac{(r_2 \dot{r}_{2o} - \dot{r}_{2o} R)^2}{v(A_2^2 + B_2^2)}} \times \sin\left(tag^{-1}(\frac{A_1}{B_1}) - tag^{-1}(\frac{A_2}{B_2})\right) \\ = \frac{r_1 \dot{r}_{1o} - \dot{r}_{1o} R}{v \sqrt{A_1^2 + B_1^2}} \quad (27)$$

Solving this equation, the value of $v$ is obtained explicitly and independently of $\theta_v$, as

$$v^2 = \frac{(r_{10} \dot{r}_{10} - \dot{r}_{10} R)^2 (A_2^2 + B_2^2)}{(A_1 B_2 - A_2 B_1)^2} + \ldots \\ \frac{(r_{20} \dot{r}_{20} - \dot{r}_{20} R)^2 (B_1 B_2 + A_1 A_2)^2}{(A_2^2 + B_2^2)(A_1 B_2 - A_2 B_1)^2} + \ldots \\ 2 \times \frac{(r_{10} \dot{r}_{10} - \dot{r}_{10} R)(r_{20} \dot{r}_{20} - \dot{r}_{20} R)(B_1 B_2 + A_1 A_2)}{(A_1 B_2 - A_2 B_1)^2} + \ldots \\ \frac{(r_{20} \dot{r}_{20} - \dot{r}_{20} R)^2}{A_2^2 + B_2^2} \quad (28)$$

Finally, substituting the value of $v$ in (25), $\theta_v$ is also obtained as

$$\theta_v = \sin^{-1}(\frac{r_{10} \dot{r}_{10} - \dot{r}_{10} R}{v \times \sqrt{A_1^2 + B_1^2}}) - tag^{-1}(\frac{A_1}{B_1}) \quad (29)$$

## References

[1] H. Yang, J. Chun, D. Chae, "Hyperbolic localization in MIMO radar systems," *IEEE Antennas Wirel. Propag. Lett.*, vol. 14, pp. 618–621, Nov. 2014.

[2] E. L. Ferguson and B. G. Ferguson, "High-precision acoustic localization of dolphin sonar click transmissions using a modified method of passive ranging by wavefront curvature," *J. Acoust. Soc. Am.*, vol. 146, no. 6, pp. 4790–4801, Dec. 2019.

[3] P. Wan, Y. Ni, B. Hao, Z. Li and Y. Zhao, "Passive localization of signal source based on wireless sensor network in the air," *Int. J. Distrib. Sens. Networks*, vol. 14, no. 3, Mar. 2018.

[4] A. G. Dempster and E. Cetin, "Interference localization for satellite navigation systems," *Proc. IEEE*, vol. 104, no. 6, pp. 1318-1326, Mar. 2016.

[5] F. Jiang, Z. Zhang, H. E. Najafabadi and Y. Yang, "Underwater TDOA/FDOA joint localisation method based on cross-ambiguity function," *IET Radar Sonar. Nav.*, vol. 14, no. 8, pp. 1256–1266, Aug. 2020.

[6] M. Nikoo and F. Behnia, "Single-site source localization using scattering data," *IET Radar Sonar Navig.*, vol. 12, no. 2, pp. 250–259, Oct. 2018.

[7] O. Bar-Shalom and A. Weiss, "Emitter geolocation using single moving receiver," *IEEE Signal Process Mag.*, vol. 105, pp. 70–83, Dec. 2014.

[8] P. E. Howland, D. Maksimiuk and G. Reitsma, "FM radio based bistatic radar," *IEE Proc., Radar Sonar Navi.*, vol. 152, no. 3, pp. 107–115, July 2005.

[9] N. Marchand, "Error distributions of best estimate of position from multiple time difference hyperbolic networks," *IRE Trans. Aerosp. Navig. Electron.*, vol. 11, no. 2, pp. 96–100, Jun. 1964.

[10] H. Lee, "A novel procedure for assessing the accuracy of hyperbolic multilateration," *IEEE Trans. Aerosp. Electron. Syst.*, vol. 11, no. 1, pp. 2–15, Jan. 1975.

[11] R. Schmidt, "A New approach to geometry of range difference location," *IEEE Trans. Aerosp. Electron. Syst.*, vol. 8, no. 6, pp. 821–835, Nov. 1972.

[12] B. T. Fang, "Simple solutions for hyperbolic and related position fixes," *IEEE Trans. Aerosp. Electron. Syst.*, vol. 26, no. 5, pp. 748–753, Sept. 1990.

[13] A. R. Zekavat and R. M. Buehrer, "Handbook of position location: theory, practice and advances," *Wiley-IEEE. Press*. Dec. 2019. Available: https://onesearch.library.uwa.edu.au

[14] W. Xu, H. Zi-Shu, "Target motion analysis in three-sensor TDoA location system," *Int. J. Inf. Technol. Manage.*, vol. 10, pp. 1150–1160, Mar. 2011.

[15] S. Jan, Y. Kao, "Radar tracking with an interacting multiple model and probabilistic data association filter for civil aviation applications," *IEEE Sens. J.*, vol. 13, no. 5, pp. 6636–6650, Mar. 2013.

[16] L. Chen, S. Li, "IMM tracking of a 3D maneuvering target with passive TDoA system," *Proc. of IEEE Conf. Neural Networks & Signal Processing.*, pp. 1611–1614, Nanjing, China, Dec. 2003.

[17] N. J. Willis and H. D. Griffiths, Advances in Bistatic Radar, SciTech Publishing Inc., Raleigh, NC, ISBN 1891121480, 2007.

[18] P. E. Howland, D. Maksimiuk and Reitsma G. "FM radio based bistatic radar," *IEE P-Radar Son Nav.*, vol. 152, no. 3, pp. 105–106, July 2005.

[19] Y. Xu, Y. S. Shmaliy, X. Chen and Y. Li, "Robust inertial navigation system/ultra wide band integrated indoor quadrotor localization employing adaptive interacting multiple model-unbiased finite impulse response/Kalman filter estimator," *Aerosp. Sci. Technol.*, vol. 98, Mar. 2020.

[20] Y. S. Shmaliy, S. Zhao and C. K. Ahn, "Unbiased FIR filtering: an iterative alternative to Kalman filtering ignoring noise and initial conditions," *IEEE Control Syst.*, vol. 37, no. 5, Oct. 2017.

[21] F. Quo and K. C. Ho, "A Quadratic constraint solution method for TDoA and FDoA localization," *n Proc. IEEE Int. Conf. Acoust. Speech, Signal Process. (ICASSP)*, May 2011, pp. 2588–2591.